\documentclass[10pt,letterpaper]{article}
\usepackage[top=0.85in,left=1.5in,footskip=0.75in,marginparwidth=2in]{geometry}

\usepackage[utf8]{inputenc}

\usepackage{cite}

\usepackage{nameref,hyperref}

\usepackage{microtype}
\DisableLigatures[f]{encoding = *, family = * }

\raggedright
\usepackage{changepage}

\usepackage[aboveskip=1pt,labelfont=bf,labelsep=period,singlelinecheck=off]{caption}

\makeatletter
\renewcommand{\@biblabel}[1]{\quad#1.}
\makeatother

\usepackage{lastpage,fancyhdr,graphicx}
\usepackage{epstopdf}
\fancyheadoffset[L]{2.25in}
\fancyfootoffset[L]{2.25in}

\usepackage{color}

\definecolor{Gray}{gray}{.25}

\usepackage{graphicx}

\usepackage{sidecap}

\usepackage{wrapfig}
\usepackage[pscoord]{eso-pic}
\usepackage[fulladjust]{marginnote}
\reversemarginpar

\begin{document}
\vspace*{0.35in}

\begin{flushleft}
{\Large
\textbf\newline{Fourth Industrial Revolution for Development: The Relevance of Cloud Federation in Healthcare Support}
}
\newline
\\
Olasupo O. Ajayi\textsuperscript{1,*},
Antoine B. Bagula\textsuperscript{1,2},
Kun Ma\textsuperscript{1}
\\
\bigskip
\bf{1} Department of Computer Science,University of the Western Cape, Bellville, South Africa
\\
\bf{2} ISAT Lab, Computer Science Department, University of the Western Cape, South Africa
\\
\bigskip
* olasupoajayi@gmail.com

\end{flushleft}

\section*{Abstract}
Inefficient healthcare is a major concern among many African nations and can be mitigated by building world-class infrastructure connecting different medical facilities for collaboration and resource sharing. Such infrastructure should support the collection and exchange of medical data among healthcare practitioners for the purpose of accessing expertise not available locally. It should be equipped with the most recent technologies of the fourth Industrial Revolution (4IR), providing decision support to doctors thereby enabling African nations leapfrog from poorly equipped to medically prepared countries. Sadly, world-class healthcare infrastructure are a missing piece in the African public health ecosystem. Medical facilities are either non-existent or prohibitively expensive when they exist. Federated cloud computing can provide a solution to this challenge. Being a model that allows collaboration between multiple Cloud Service Providers (CSPs) by pooling computing resources together with the aim of meeting specific business or technological need; it allows for the execution of tasks on computing resources in a flexible and cost efficient manner. This paper aims to connect unconnected medical facilities in Africa by proposing a cloud federation for healthcare using co-operative and competitive collaboration models. Simulations were carried out to test the efficacy of these models using five different workload allocation schemes: First-Fit-Descending (FFD), Best-Fit-Descending (BFD) and Binary-Search-Best-Fit (BSBF); Genetic Algorithm meta-heuristic and the Stable Roommate Allocation (SRA) economic model for both light and heavy workloads. Results of simulations revealed that the co-operative cloud federation model resulted in lower allocation delays but higher resource utilisation; while the competitive model provided faster service delivery and better Quality of service (QoS) adherence. It also showed that BSBF and BFD gave the best resources utilisation and energy conservation, while FFD was the fastest overall. Finally, deployment considerations and potential business models for the federated cloud for healthcare in Africa were presented.


\section*{Introduction}
Cloud computing is a key technology which plays a vital role when interfacing the physical and virtual worlds in most fields of the fourth industrial revolution (4IR). There are numerous definitions of Cloud computing in literature, however that of the NIST is arguably the most accepted. According to the NIST, Cloud computing is a model that enables pools of measurable computing resources be made available to users conveniently and ubiquitously \cite{Mel11}. One of the key characteristic of Cloud computing is elastic pool of resources, this implies a near infinitive resource scale. In actuality however, no Cloud Service Provider (CSP) is able to provide a limitless amount of resources to users. Beyond elasticity, Cloud resources need to be available at any time and from any location, globally. Although it is possible to achieve global coverage for a single site data centre, users would however experience increased latency/delay and reduction in throughput as distance grows. To this end, CSPs often have data centres located in multiple geographical areas in order to be as close to the users as possible - a concept known as multi-homing \cite{mh}. In the same vein, there are situations whereby a CSP does not have enough resource to cater for all its users; such a situation might arise for example during peak office hours (company websites), during promotions and sales (for e-commerce websites) or when students are resuming new academic sessions (for academic websites). Two potential solutions to problem of resource shortage and/or extended coverage are resource scaling (either vertically or horizontally \cite{scale}) and collaboration with other CSP. Resource scaling might however be extreme costly especially if demand spike is only for a limited and short amount of time. CSP collaboration on the other hand might prove to be a more cost-effective solution.

Cloud federation has emerged as a solution for CSP collaboration \cite{cloFed}. It is based on the economic model of federation game and one in which multiple CSPs combine their resources, in a way that allows for cross-utilization amongst themselves and improve quality of services (QoS) rendered to users \cite{darz1}. Cloud federation also provides CSPs with an extended reach, as they are able to leverage on partner CSPs to reach disperse geographical locations. Cloud federation can be provided in one of three models \cite{darz1, darz2}, which are: infrastructure pooling (where resources of multiple CSPs are aggregated together and appear as a single virtual infrastructure, much similar to the disk striping or RAID0); hybrid federation, which combines resources across private and public Clouds and broker-based federation, wherein each CSP remains independent but are conjoined by a single broker. The focus of this paper is on the third model and one in which CSPs have the option to choose to join a federation or work independently. 

\subsection*{Cloud Federation for Healthcare Support in Africa}
\begin{figure}[h]
\centering
\includegraphics[scale=0.5]{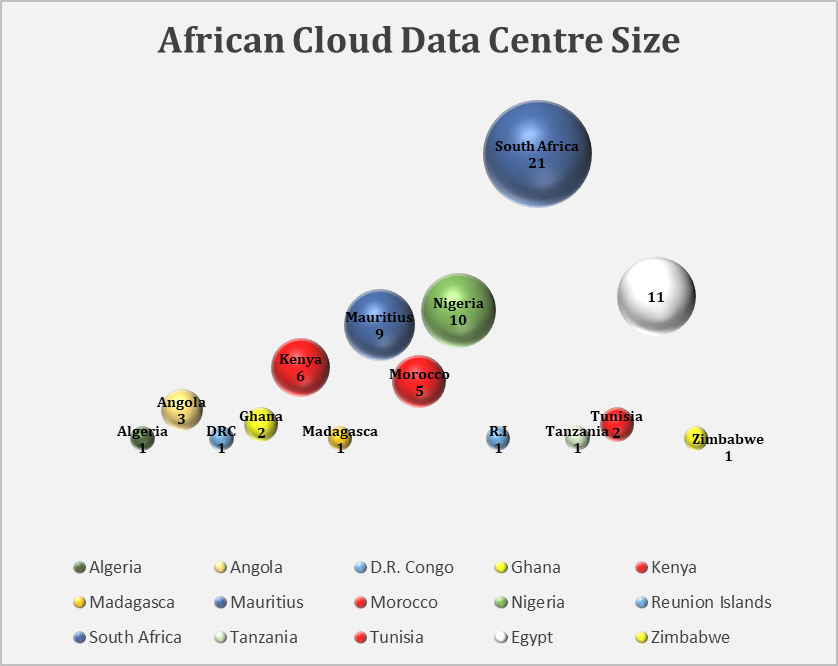}
\caption{Sizes of Data Centres across Africa}
\label{dcAfr}
\end{figure}

It is widely recognised that developing nations have missed many of the opportunities offered by the first three industrial revolutions. It is also expected that, cognisant of this sad fact, many developing countries will take advantage of the technologies of the fourth industrial revolution which are most relevant to their needs to leapfrog from poorly equipped to technologically prepared countries. The specific 4IR use-case scenario being considered in this paper is the application of Cloud federation to health care and medicine across African countries. This would allow for collaboration and resource pooling across the continent for improved health care services. The justification for a federated Cloud for medicine in Africa are numerous, few amongst which are: i. most African countries are either under-developed or developing. ii. access to world-class medical services is either non-existence or extremely expensive; however there are a handful of African countries with  good medical facilities, which can offer tele / cyber-health \cite{tmed} supports. iii. patients in many developing parts of Africans cannot afford the huge cost of flying abroad or to other African countries such as South Africa and Egypt for treatment. Cloud federation can therefore allow collaboration, wherein resources can be pooled together to carry out processes such as X-Rays and CT Scans interpretations, remote testing and diagnosis, and possibly conference surgeries - where multiple experts monitor and observe surgical procedures. 
To put this in context, we would describe an application scenario. Currently, there are only about 75 Cloud data centres (DC) across the African countries according to \cite{dcmap} and Fig. \ref{dcAfr} shows their distribution, with each bubble sized proportionality to the number of DCs in each country.

\begin{figure}[h]
\centering
\includegraphics[scale=0.5]{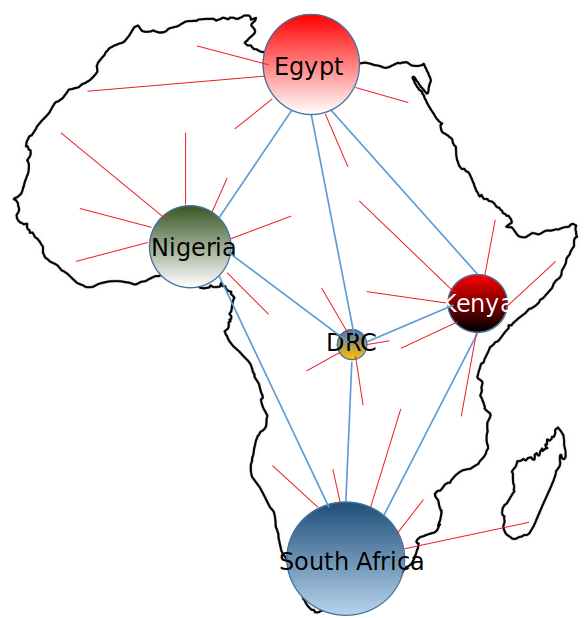}
\caption{High-level Conceptual Cloud Federation Network for Health care in Africa}
\label{spread}
\end{figure}

From Fig. \ref{dcAfr}, only six countries have more than five DCs, while nine countries have between 1-3 data centres. This is a total of fifteen countries of the total fifty-four in Africa. The other countries either do not have or theirs' are below the DC standards as stipulated in \cite{uptime,TIA}.  Building DCs and capacity is a very expensive and time consuming process. Many African countries are still encumbered with economic sustainability and survival challenges to be considering Cloud DCs. Cloud federation can therefore be of immense value to these countries and the African continent in general. Fig. \ref{spread}, shows a potential high-level Cloud federation network for medicine across Africa. Countries with a multiple DCs are chosen as regional hubs and distributed as follows: Egypt to the North, South Africa to the South, Kenya to the East, Nigeria to the West and DRC at the centre. A high bandwidth, low latency network connection between these hub nations would serve as the backbone of the federated system, while the hub countries serve as regional gateways into the network. 

The Federation can be done in one of two models. In the first, the CSPs agree to work together, forming a single virtualized resource pool; we refer to this model as a co-operative federation. On the other hand the CSPs can decide at specific time intervals to work independently, we refer to this model as the competitive federation. 

\subsection*{Contribution and Outline}

For this work, we also consider five different workload allocations schemes to determine their effects on the co-operative or competitive Cloud federation. These allocation schemes include the heuristic models - First-Fit Descending, Best-Fit Descending and Binary-Search-Best-Fit; meta-heuristic model - Genetic Algorithm and an economic model in Stable Roommate Allocation. Resource utilization, QoS and allocation delays were considered as performance metrics.  
The specific contributions of this paper are:
\begin{itemize}
\item A detailed performance comparison of five different workload allocation schemes and how they affect various metrics in co-operative and competitive cloud federations.
\item A unique GA gene encoding scheme for the allocation of Cloud workloads to PMs.
\item Potential business models for deployment of federated Cloud for health care in Africa.
\end{itemize}

The rest of this paper is organized as follows: following this introduction is a review of related work in section 2. In section 3, the Cloud federation models are presented, while the various workload allocation schemes considered are presented in section 4. Results of simulations done are presented and discussed in section 5. Deployment considerations and potential business models are presented in section 6; while section 7 concludes the paper with motivations for future works given.

\section*{Related Works}
With respect to collaboration across nations, a number of solutions already exist particularly in the academic and research domain. One such is the African Research and Education Network (AfREN), which is a network established for collaboration and research across Universities and research centres across Africa \cite{afren}. It is a region based network and consists of ASREN covering the North and Mid-East Africa, WACREN for the West and Central Africa and UbuntuNet for the East and South African countries. Similar networks also exists globally such as the Asia-Pacific Advanced Network (APAN) \cite{apan}, GEANT \cite{gnt} in Europe and internet2 \cite{int2} in the USA.

A number of works have been done on providing infrastructure to support health across the African continent. In the work done by Bagula, et al. \cite{hframe}, the authors proposed a multi-layered framework for Cyber-Physical Healthcare which combines IoT and machine learning techniques. IoT was used for collection and muling of health data, while the machine learning techniques were used for patient triage. The potential advantage of this framework include better patient prioritization, better patient monitoring as well as cost and time savings. In a related work on IoT and healthcare, considerations for designing a full stack Remote Patient Monitoring system (RPM) for tele-medicine based on FiWARE was presented in \cite{fiware}. FiWRE advocates openness and the authors proposed a solution inline with the FICHe guidelines. Critical considered to note when building such a system were give, some of which included design steps, device deployment, data - collection, muling, security and storage as well as system integration.

 The authors in \cite{mdsa}, considered the applications of Fog computing for storing sensitive health information in a Cyber-healthcare system  in a secure manner and proposed the Multi-Phased Data Security and Availability (MDSA) protocol. The fog networks helped cut down network latency, while the multi-phased security ensured end-to-end security coverage. In another related work, the authors in \cite{priority}, proposed a Cloud-based medical triage service system. Upon collecting body vital signs from patients, the system analyses the information using either linear regression or k-means, benchmarking the obtained results again the WHO standard. 
In order to achieve collaborative health care system pan-African wide, standards have to be agreed upon for effective transmission and interpretation of patient medical / health records. This would foster interoperable between the various Cloud platforms spread across the continent. Lubamba and Bagula \cite{hl7cda}, had proposed a framework for the standardization of medical data. Their proposed model was based on the Health Level Seven (HL7) standard \cite{hl7s}. In their work, patient data had to be encoded into XML based HL7 format before being transmitted using HL7-CDA web service. From obtained results, the authors showed that their HL7 based model was able to transmit significantly more records, with minimal overhead impact when compared with the alternatives.  The works above could be applied in the implementation of healthcare kiosks in developing countries as suggested in~\cite{kiosks,cyberhealth}. 

In the work done by Shimizu \emph{et al.} \cite{apan}, the authors presented medical use cases of combining the Asia-Pacific Research and Education Network (REN) with a Digital Video Transport System (DVTS). The DVTS allowed them obtain digital streams of images which could be transported via an IP network, while the REN provided a stable high-bandwidth network for transmission. A hundred different medical teleconferences were used as test, with images from live surgical sessions, endoscopy, transplants, nursing and health care amongst others. The authors in \cite{mars} also discussed on the potentials and advantages of introducing Tele-medicine in Africa, some of which includes: lowering medical costs, reducing geographical distance and cater for severe shortage of doctors across the African continent. Factors limiting the wide-spread adaptation of Tele-medicine as well as possible future directions were also presented.

Cloud computing has in the last two decades emerged as a reliable, robust and capable computing paradigm. It has therefore seen numerous practical application in almost all aspects of live. Cloud computing has also grown beyond the single-site, single provider solution it once was to one in which multiple CSPs work together to achieve set goals. Darzanos, \emph{et al.} \cite{darz1}, had proposed a model for economically evaluating Cloud federation. They focused on workload delays within federated Cloud systems. With time being the main metric, they therefore modelled each CSP as an M/M/1. In the work, the resources of the CSPs were pooled and user workloads could be served by resources belonging to any of the participating CSPs. They finally developed a model for allocation that maximized the profit of the collective whole. In a latter work \cite{darz2}, the authors still considered each CSP as an M/M/1 but also considered the performance across three types of cloud federation models- weak, strong and elastic. The strong is a completely co-operative model, the weak is similar a competitive model, while the elastic can be described as a dynamically competitive model. In this work however, profit was dependent on energy consumption and QoS-adherence. Finally, the Shapley-value was used to profit sharing among the participating CSPs.

With respect to our choice of allocation schemes, we can consider the allocation of workloads to Cloud resources as a bin-packing problem \cite{bin}, which in itself is a NP-hard problem. This therefore necessitates the use of non-intrinsic methods to solve, such as heuristic and meta-heuristic models. In terms of the heuristic, the first-fit, best-fit and their variance are arguably the most common. For Cloud workload allocation, the Best-fit-descending (BFD) has been widely used by numerous researchers \cite{BelBuy, Ban13, Hieu17}, it therefore makes an excellent choice for our selection. First-fit-descending like BFD uses has been shown to use the same amount of bins, but FFD is faster. With allocation speed being a core metric in this paper, we therefore considered FFD. In terms of meta-heuristic, the Genetic Algorithm has been widely used in many literature for workload allocation in Cloud computing environs. A few of these works are \cite{dah17, Liu22, oshin}. For a number of these work, energy conservation, QoS and resource utilization were metrics considered.   

Economic concepts have been widely applied in solving computing related problems. Coalition games and game theories have been used for problems relating and involving multiple participants - such as in VM migration in federated computing \cite{anchor14} and dynamic resource re-allocation in \cite{anchor17}. The stable marriage/roommate economic models have been also used for workload allocation in Cloud computing environments \cite{anchor}.
 
The focus of a number of these works were either on medical collaboration via the Internet and/or various schemes for allocating workloads to Cloud resources. Unlike in these other works, in this paper we consider a Cloud federation system for improved user satisfaction and service delivery across the African continent. Like the work of \cite{darz2}, this work also compares the different cloud federation model, however we seeks to determine which of the two models is best suited for specific requirements - light or heavy usage demands. To achieve this, we applied various workload using a simulated federated network consisting of multiple CSPs working co-operatively or competitively to provide medical services across the continent. 

\section*{The Cloud Federation Model}
Typically, a  federated model for Cloud computing includes different cloud providers collaborating by: i) sharing their resources while having each of them remaining an independent autonomous entity by keeping "thick walls" in between them; ii) having the applications running in this cloud of clouds while being unaware of location due to virtual local networks being designed and implemented to enable  the inter-application components to communicate and iii)  having cloud providers differentiate from each other in terms of cost and trust level.

\subsection*{Cloud Federation Model Formulation}
\begin{figure}[h]
\centering
\includegraphics[scale=0.25]{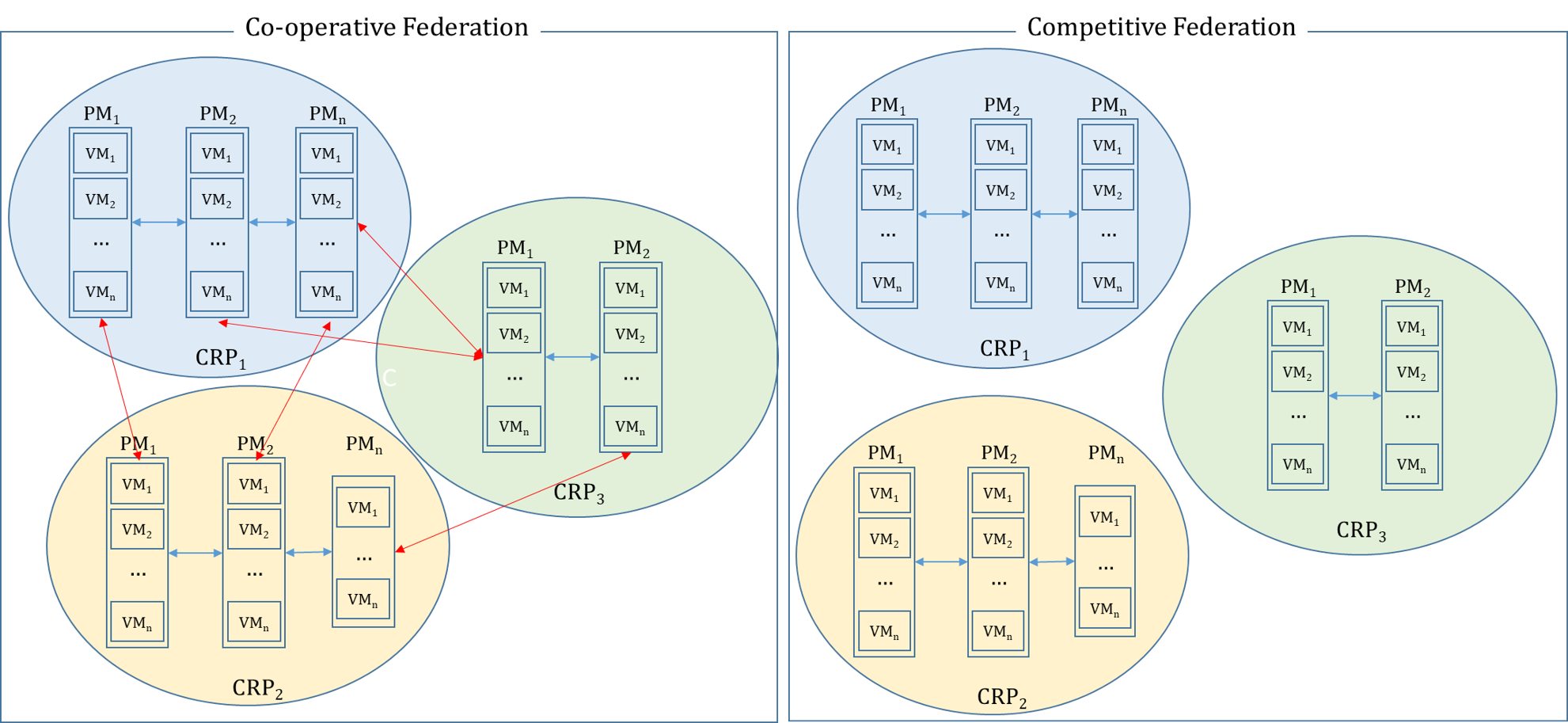}
\caption{Co-operative vs Competitive Cloud Federation Models}
\label{coopComp}
\end{figure}

When considering a federated cloud environment, the virtual machines allocated to the users' tasks can be migrated either to physical resources of the users' cloud provider or to physical resources of different cloud providers. Such an allocation of virtual resources to physical resources can lead to a cooperative model when users' virtual machines can be migrated anywhere or a competitive model when users' virtual machines can only be migrated to their providers' physical machines as expressed by Fig. \ref{coopComp}.

In our model, the resources which have been availed by  a  cloud provider $j$ are expressed by $R_{pn}(j)$ while demand for resources by a VM $i$ during migration are expressed by $D_{vn}(i)$.  The federated cloud computing problem consists of finding for each VM $i$ in distress, a mapping to a physical resource provider $j$ that maximises a utility function $D(i,j)$ as defined below

\begin{equation} \label{diff}
\begin{array}{ll}
\max D(i,j) & = \alpha(i,j) * (R_{pn}(j) - D_{vn}(i))   \\
subject & \mbox{ to } 
\end{array}
\end{equation}

$$
\left \{
\begin{array} {llr}
R_{pn} (i) \ge D_{vn} (j)  	& \forall i \in {\mathcal V}, j \in {\mathcal P}    &\mbox{     (\ref{diff}.a)}\\
\alpha \in \{0,1\} 		& \forall i \in {\mathcal V}, j \in {\mathcal P}    &\mbox{     (\ref{diff}.b)}
\end{array}\right.
$$
Note that as expressed by equation $\ref{diff}.b$,   $P(i,j)$ is a binary parameter used in the model to differentiate between cooperative and competitive cloud computing as expressed below:
\begin{equation} \label{pij}
\alpha(i,j) = \left \{
\begin{array} {llr}
1   & VM(i) \in PM(j)       &  \mbox{   Cooperation}\\
0   & VM(i) \notin PM(j)  & \mbox{    Competition}
\end{array}\right.
\end{equation}
Note that as expressed by equation~\ref{pij},  $P(i,j)$ is used in the model to enable all participant providers to be elected for a VM migration under cooperative cloud computing and discard providers from participating in a Vm migration under competitive cloud computing when the VMs don't belong to their clients. 

\subsection*{Cloud Migration Process}

\begin{figure}[h]
\centering
\includegraphics[scale=0.4]{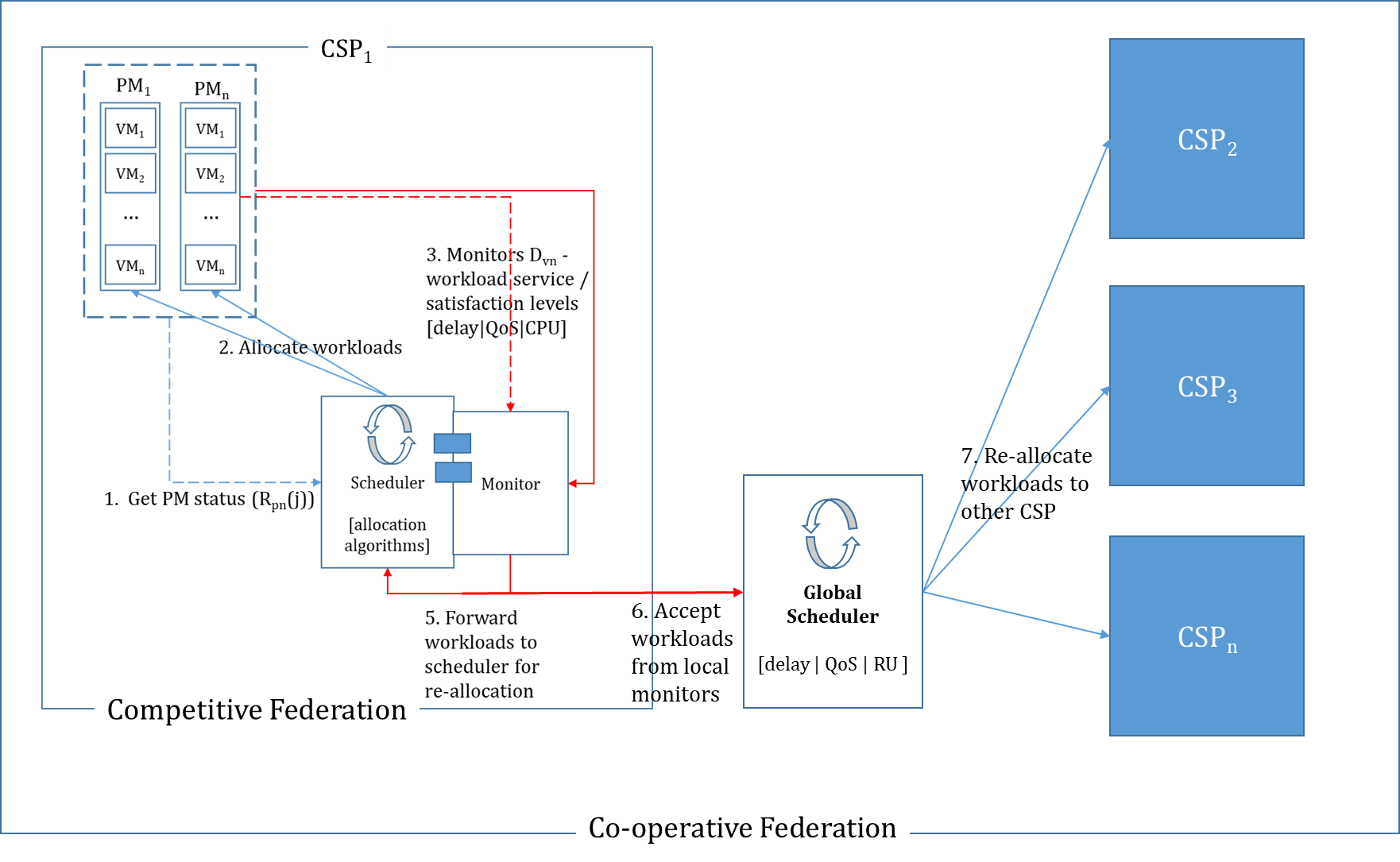}
\caption{Migration Model for Federated Cloud}
\label{fedMig}
\end{figure}

Fig. \ref{fedMig} shows the processes involved in workload migration in federated Cloud. The process starts with the allocation of workload to physical machine. The allocation is done in a way that the size of the Cloud resource $j$ $(R_{pn}(j))$ meets or exceeds the workload $i$'s requirement $(D_{vn}(i))$. With continuous allocation, the cloud resource $j$ might becomes unable to meet workload requirements equation~\ref{diff}, hence the need to migrate workloads to other viable resources.
The monitor - a component of the scheduler handles this process. 

In the competitive federation, the workloads selected for migration are forwarded to the scheduler for re-allocation to other resources. This process is depicted in CSP$_1$ of Fig. \ref{fedMig}. In the co-operative however, the selected workloads are forwarded to the global scheduler for re-allocation into a different cloud resource in the same or different CSP. This is as illustrated on the right of Fig. \ref{fedMig}.

\section*{Cloud Workload Allocation Models}
We have described the allocation of workloads to Cloud resources as a bin packing problem. In this section, we present the five various allocation models considered in this work for "packing" workloads into servers. The models are described as follows:

\begin{enumerate}
   \item \textbf{Best-Fit Descending (BFD):} BFD is a greedy heuristic algorithm that has been shown to use \((11/9 * optimalBins) + 1 bins\) \cite{bin}. When applied in Cloud computing, virtual machines (VMs) are considered items to be put in bins while the physical machines (PMs) are considered as the bins. Both the VMs and PMs are of heterogeneous sizes. The allocation speed of BFD can be increased if the PMs are sorted in order of their capacity. In this work only the CPU is considered, thus the PMs are sorted in decreasing order of CPU. This is done to allow for a uniform basis of comparison across all the different workload allocation models compared.
  \item \textbf{First-Fit Descending (FFD):} FFD is another variant of the greedy heuristic algorithm but unlike BFD, it assigns VMs to the first PM it finds that can accommodate it. The performance of this algorithm can also be significantly improved if the PMs are sorted in descending order, thus reducing search time for suitable PMs.
\item \textbf{Binary-Search Best-Fit (BSBF):} This is an algorithm proposed in \cite{BSBF}, with the main objective of speeding up the PM search time. Rather than the linear PM search used by BFD and FFD, it instead builds a Red-Black Tree (RBT) using PM capacity (available CPU).  Being a RBT, in theory it has a worse case search time complexity of \(log_2n\) which is faster than BFD and FFD with complexities of at least n. There is however an additional time required to build and update the RBT which also needs to be taken into consideration. This notwithstanding, BSBF was reported by the authors to still be significantly faster than the other fit algorithms and conserves resources better; hence the reason for considering it in this paper.
\item \textbf{Stable Roommate Allocation (SRA):} For this work, the stable roommate algorithm was adapted for application in Cloud workload allocations. The stable roommate is a version of stable marriage wherein one party is allowed to have multiple partners or a room is allowed to have more than one occupant. PMs are taken to represent men/rooms while VMs represent the user workloads to be allocated. Multiple VMs can be assigned to a PM, but a VM can only be assigned to a single PM. In implementing this, and similar to the work of Xu and Li \cite{anchor}, the VM preference list was built by considering PMs with available CPU greater than the VM's, while PM preference is based on VMs requests that a less than the PM's available CPU capacity. Prior to allocation, each room / man proposes to VMs. The VMs do not immediate accept the proposal(s) but adds them to a list of suitors. During the allocation phase, each VM cross references its list of suitors with its preference list and only accepts proposals that best suits it. The process is repeated until all VMs are matched to PMs. In this work, priority is placed on ensuring that all VMs are allocated. Hence it is possible for some PM(s) not to be matched to any VM. In fact this is desirable as it implies better resource utilization and lower energy consumption. Fig. \ref{sm} shows an illustration with four PMs, \(p_1...p_4\) and five VMs, \(v_1...v_5\). Both PMs and VMs have their preference list, while each VM also has a suitor list (list of PMs that proposed to it).

\begin{figure}[h]
\centering
\includegraphics[scale=0.25]{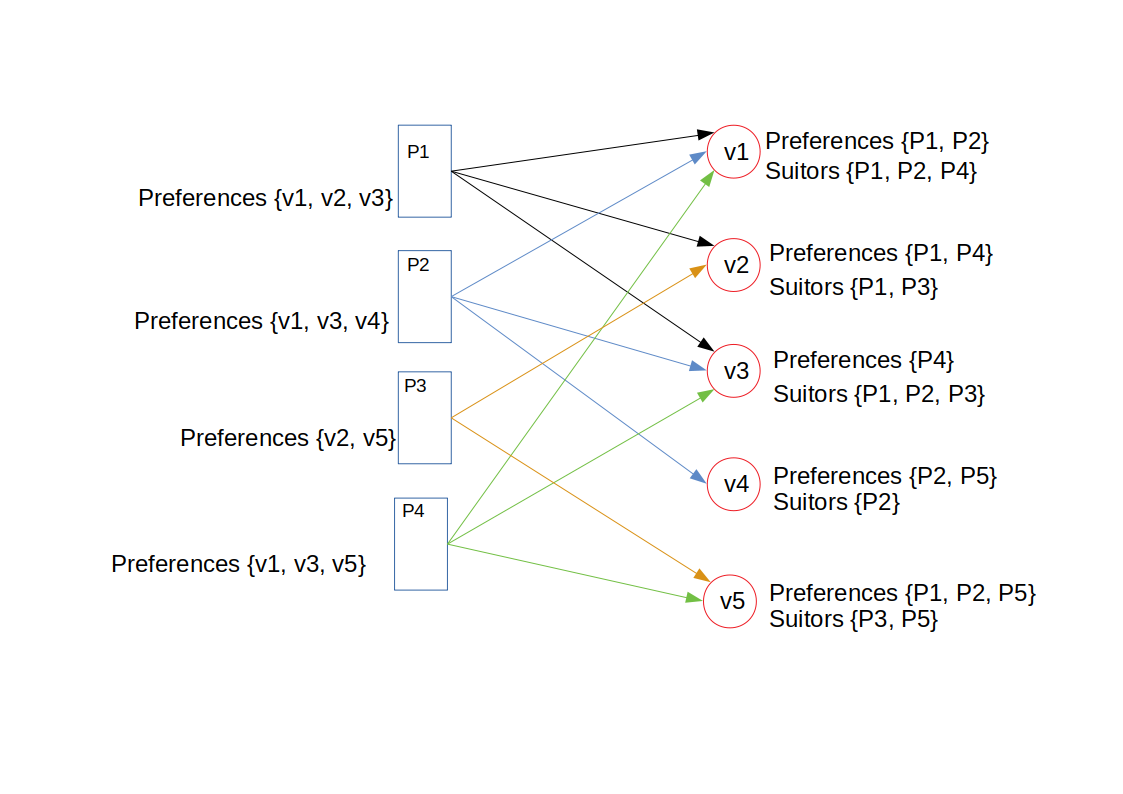}
\caption{Illustration of Stable Roommate Allocation}
\label{sm}
\end{figure}

\item \textbf{Genetic Algorithm-based VM Allocation (GAVA):} There have been a number of works that have applied GA to Cloud resource allocation; of particular note are \cite{mosa, oshin}. Like in those works, we also followed the classical GA steps described in \cite{chen12}, however our implementation was a bit different. We assumed a PM to be made up of two processing elements (PE). We then took a PM to represent 2 genes and a string of genes (chromosome) to represent a potential VM-PM allocation solution. In encoding our genes, when a PE is potentially allocated to a VM, we set it to 1 and to 0 if otherwise. This is as illustrated in Fig. \ref{gene}. We performed partial mutation and only changed 0s to 1s. This was because changing a 1 to 0 would require de-allocating all VMs currently assigned to such a PE and then looking for new potential VMs to allocate. Rather than performing a repeated allocations, we created a different chromosome instead. Finally, we took fitness value to imply the number of 0s in the chromosome. Therefore, the chromosome with the highest number of 0s was selected as the best. This translates to a solution which uses the least amount of PMs to serve all VMs.

\begin{figure}[h]
\centering
\includegraphics[scale=0.25]{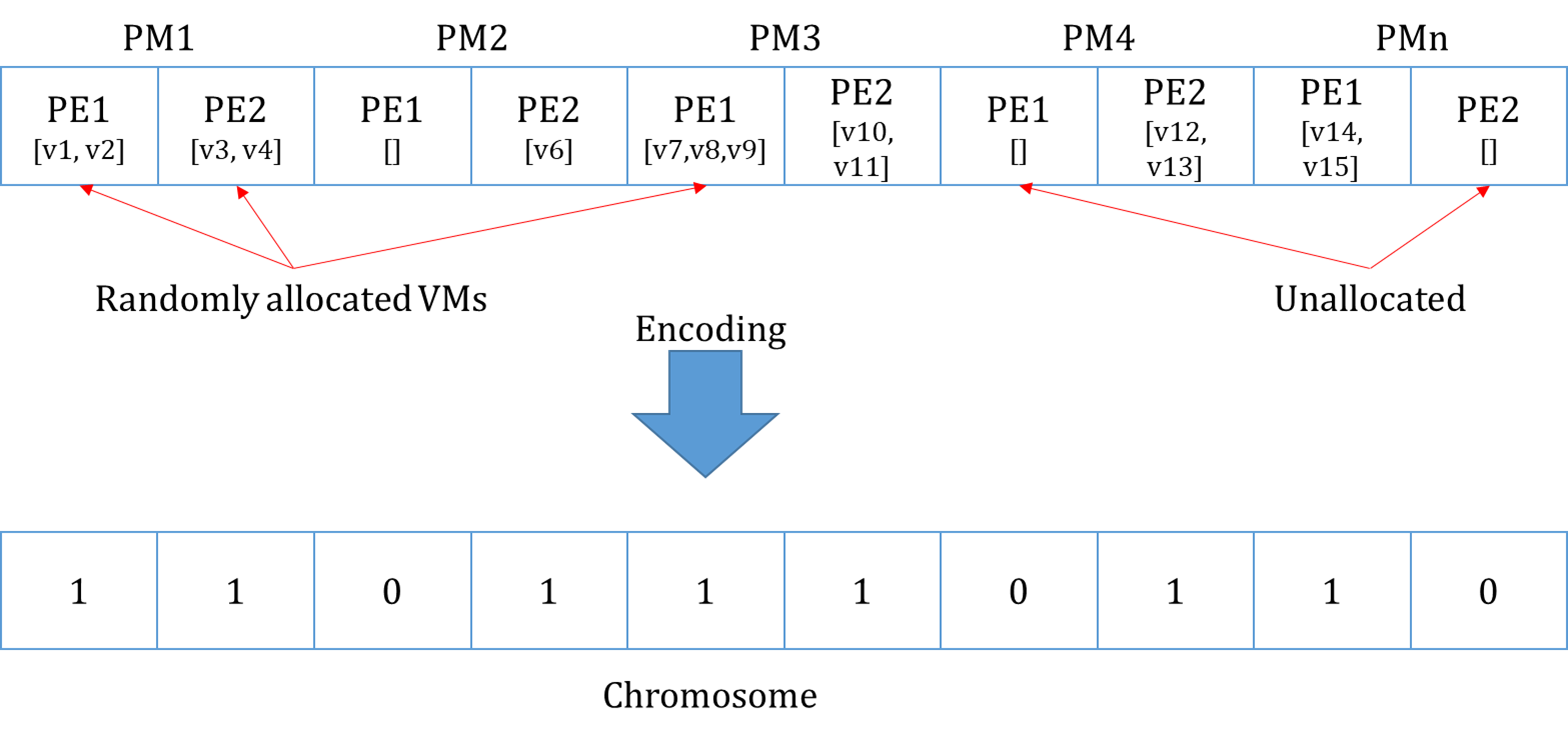}
\caption{GA Gene Encoding}
\label{gene}
\end{figure}

\end{enumerate}

\section*{Results and Discussion}

For this work, simulation were carried out using Cloudsim \cite{clo} and a data centre similar to that used in \cite{BelBuy, BSBF, mosa, Hieu17} was used and consisted of a number of heterogeneous PMs. These PMs were of two categories with specifications and power consumption models based on benchmarked data from real servers \cite{rice15} and given as follows: category one had 2 CPU cores clocked at 1,860MHz and 4GB of memory, while the second category also had 2 CPU cores each clocked at 2,600MHz and with 4GB of memory. 

To model the co-operative federated Cloud: a data centre with a total of 300 PMs was setup in CloudSim. User workloads were executed on any of four types of VMs, viz.: single core @ 2500MHz, single core @ 2000MHz, single core @ 1000MHz and single core @ 500MHz. Data used for this experiment were from anonymized workload traces of VMs submitted to a Google cluster and PlanetLab. A total of 168 workload traces were used for each experiment and distributed as follows:
\begin{enumerate}
    \item To simulate light user demands, the smallest 56 traces from the Google cluster TraceVersion1 \cite{trace1} were used. 
    \item For heavy demands, 56 of the largest traces were extracted from PlanetLab dataset of 12th April, 2011 \cite{plab} and used. 
    \item For the medium, the 56 traces used were made up of a mix of large traces from Google cluster and light traces from PlanetLab dataset.
\end{enumerate}

For the competitive federation on the other hand, three data centres were set up to simulate the countries with the most number of DCs as shown in Fig. \ref{dcAfr}. For a fair and consistent result, we assigned equal number of PMs to the countries, at 100 each. Similar to the co-operative, user workloads were also split into light, medium and heavy and were ran on VMs with similar configuration as those used for the co-operative federation.

In presenting the results, the performance of both federation models under light and heavy workloads were compared, those of medium workloads were omitted in order to conserve space. Seven metrics were considered and the obtained results are presented in the subsequent subsections:

\subsection*{Light Workload}

\subsubsection*{Allocation Delay}
This is a measure of how long users have to wait before processing begins on their submitted workloads. Two delays are considered in this work, pre-processing delay and average delay. 

\begin{enumerate}
\item \textbf{Pre-processing Delay:}
The pre-processing delay is a measure of the time spent by each algorithm before allocating the first user workload (VM). For BFD and FFD, it is the time spent sorting all PMs in descending order of available CPU. For BSBF, it is the time spent sorting the PMs in descending plus time spent on building the binary search tree. For GAVA, pre-processing delay, was time spent encoding genes, building up a population of chromosomes and iterating through 200 generations to find the best individual (VM-PM mapping); while for SRA, it is the time spent building the preference list for both PMs and VMs as well as the time it took each PM to propose to all its preferred VMs. Only the pre-processing times of BSBF, BFD and FFD are reported, this is because GAVA and SRA had their pre-processing done offline as they took significantly longer time to complete compared to the others. The results of pre-processing times are shown in Fig \ref{fallo}.

From the figure, the algorithms had varied pre-processing times under the two federation models. BSBF has the longest pre-processing delay for both the co-operative and competitive models at 3,943,250ns and 4,253,600ns respectively. This is due to the extra time spent building the binary search tree. BFD was second, at 1,795,000ns for co-operative and 3,533,900ns for competitive. FFD was the fastest of the three at 509,300ns for co-operative federation and 785,833ns for the competitive. Cumulatively, pre-processing delays were higher in competitive than in the co-operative federation. 

\begin{figure}[h]
\centering
\includegraphics[scale=0.5]{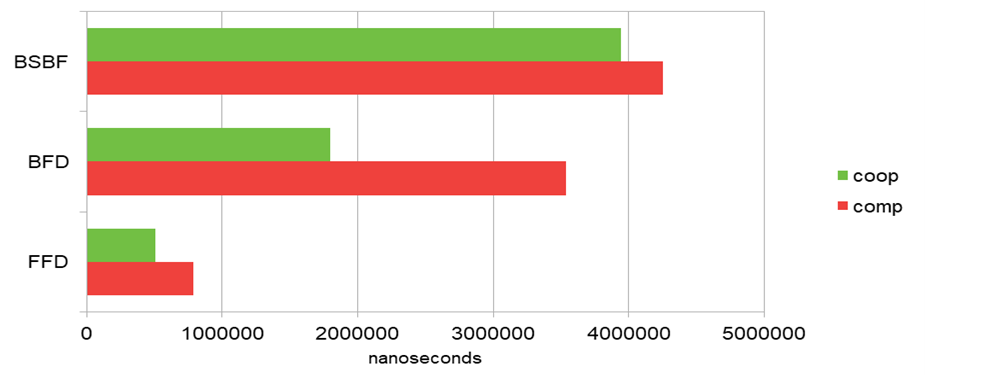}
\caption{Pre-processing Delay}
\label{fallo}
\end{figure}

\item \textbf{Average Delay:}
A measure of the average time taken to allocate a VM to a PM. The results of this is shown in Fig. \ref{avg}. BFD took the longest time, across both federation models, at 756,459.00ns for co-operative and 1,174,501.67ns for competitive. Conversely, FFD reported the least allocation delay. For both federation models, FFD and BSBF, gave almost equal delays with the competitive only marginally faster (less than 3,000ns) in both cases.

\begin{figure}[h]
\centering
\includegraphics[scale=0.5]{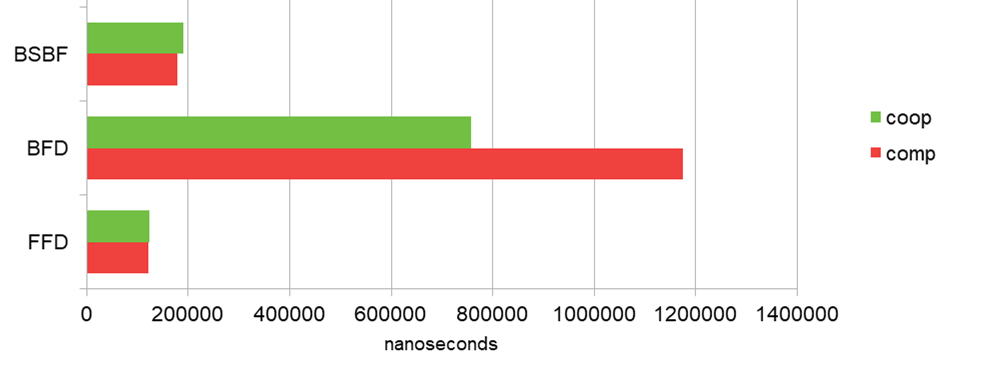}
\caption{Average Workload Allocation Delay }
\label{avg}
\end{figure}

This significant difference in speed between the algorithms can be attributed to their mode of operation. BFD searches through the entire list of PMs for one that best fits a given workload, while FFD assigns to the first capable PM it finds. The benefit of the binary search tree used by BSBF is most evident here and responsible for the lower delays compared to BFD and almost as fast as FFD. This observation is in line with that reported in \cite{BSBF}.  

\end{enumerate}

\subsubsection*{Execution Time}
\begin{figure}[h]
\centering
\includegraphics[scale=0.5]{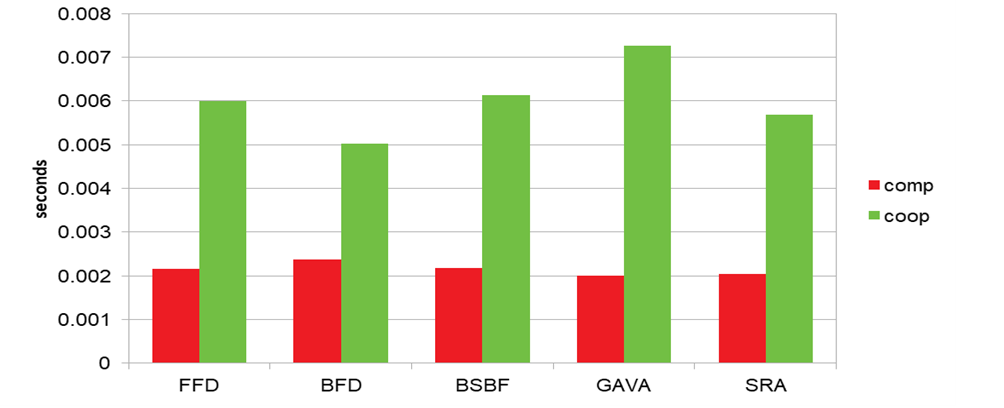}
\caption{Comparison of Overall Completion Times}
\label{time}
\end{figure}

In this paper, execution time is taken to mean the total time spent servicing user workloads. Fig. \ref{time} shows a comparison of execution times of the different allocation schemes for both federation models when light workloads are submitted. For the co-operative federation, BFD resulted in the quickest execution time, this is followed by SRA, FFD, BSBF and finally GAVA. For the competitive federation, SRA and GAVA were the quickest, followed by FFD, BSBF and BFD. It is important to note that these times difference are only in factions of seconds. For all algorithms, workload execution took shorter time to complete in the competitive federation than in the co-operative. 

\subsubsection*{Resource Utilization}
\begin{figure}[h]
\centering
\includegraphics[scale=0.5]{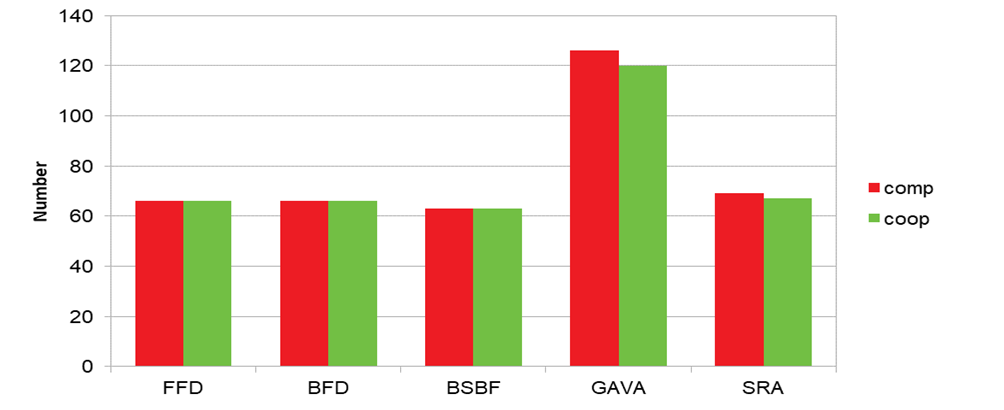}
\caption{Comparison of Resource Utilization}
\label{rub4}
\end{figure}

The five allocation algorithms were compared to determine how well they utilize resource when allocating workload to PMs. Two results are presented, the first being resource utilization right after allocation and the second being after optimizing the allocation. Optimizing the allocation aims to reduce the number of resources used by consolidating workloads into fewer number of PMs. From the result shown in Figure \ref{rub4}, for both co-operative and competitive federation equal number of resources were used across all but GAVA and SRA. For GAVA, co-operative federation was slightly better with 120 PMs versus 123 in competitive federation; similar results were obtained for SRA with 67 PM for co-operative and 69 for competitive. Across all allocation, BSBF resulted in the best matching of VMs to PMs and utilized only 63 PMs. BFD and FFD followed closely with 66 PMs, while GAVA gave the worst. It must however be noted that, the result is based on a fitness function set to 65\% utilization and 200 epochs; a lower fitness function and more epochs might have resulted in lowered values, though at the cost of an even longer training time. Cumulatively, the co-operative federation model was slightly better as it utilized an average of 76 units of resource compared to the 78 used in the competitive federation.  

\begin{figure}[h]
\centering
\includegraphics[scale=0.5]{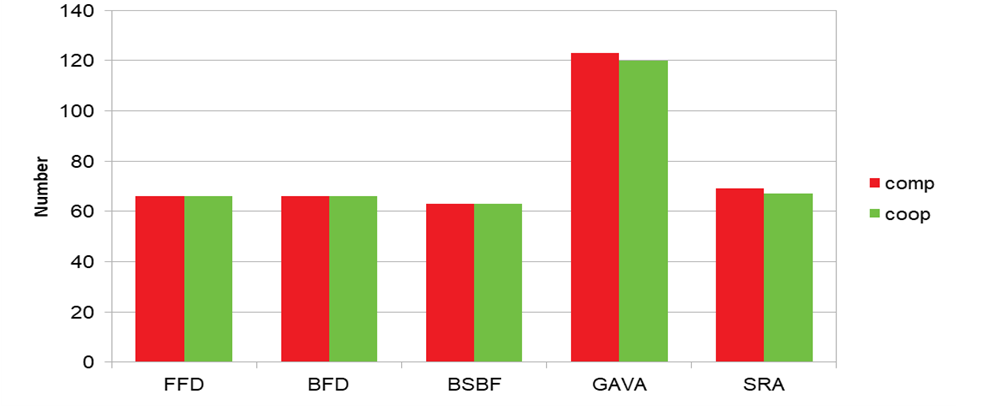}
\caption{Resource Utilization after Consolidation}
\label{ruafter}
\end{figure}

Fig. \ref{ruafter} shows the utilization after workload consolidation. From the results and across all algorithms, the co-operative federation was again better than the competitive but only marginally at an average of 76 and 77 resource units respectively.

The main purpose for considering the resource utilization after consolidation is to determine how well each algorithm performed in terms of packing workloads into PMs. The lower the change in number of resources utilized between "before consolidation" and "after consolidation", the better the algorithms is at packing. 

\subsubsection*{Energy Conservation}
\begin{figure}[h]
\centering
\includegraphics[scale=0.5]{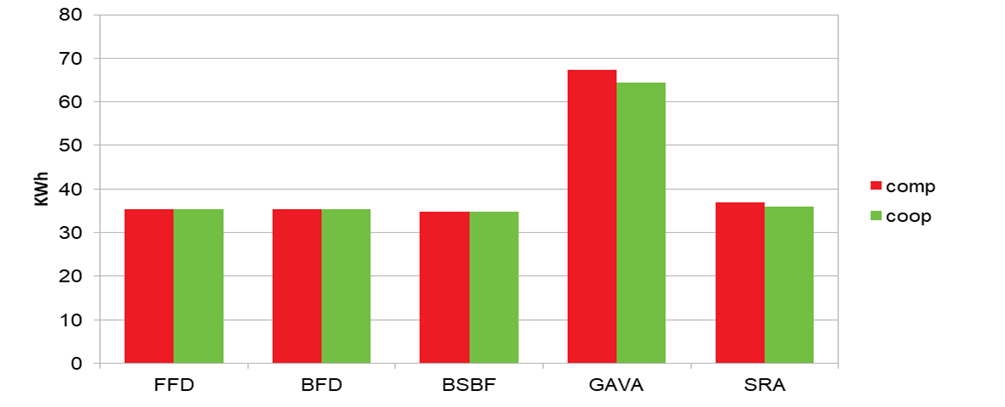}
\caption{Energy consumption before consolidation}
\label{energy1}
\end{figure}

\begin{figure}[h]
\centering
\includegraphics[scale=0.5]{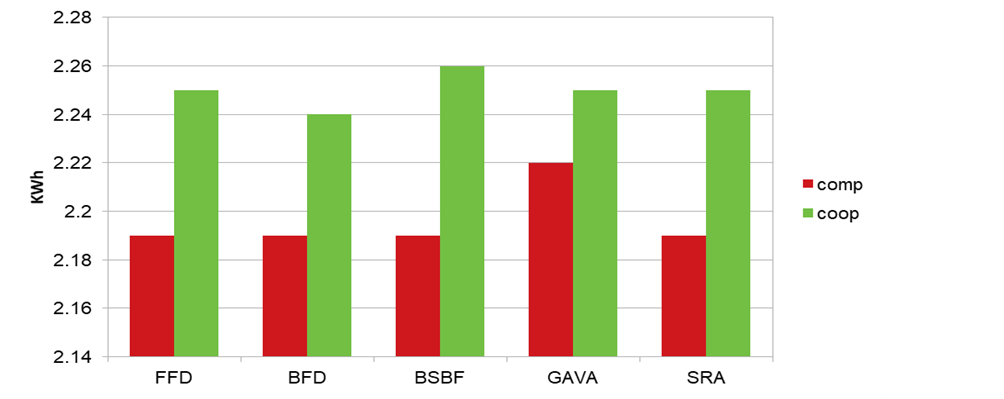}
\caption{Energy consumption after consolidation}
\label{energy2}
\end{figure}

Besides effective resource utilization, conservation of energy is also very vital to CSPs, as there is a global drive to reduce energy consumption and carbon emissions for the purpose of a greener earth. Comparisons of the five algorithm with respect to energy conservation for both federation models are shown in Fig. \ref{energy1} and Fig. \ref{energy2}.

In Fig. \ref{energy1}, energy consumption levels were almost similar across all algorithms and for both federation model. This is in line with the resource utilization levels shown in Fig. \ref{rub4}. Overall, energy consumption in co-operative federation, were slightly better than that of competitive. Furthermore, BSBF with 34.8KWh conserved energy the most across both federation models. It was followed by BFD and FFD both at 35.4KWh; SRA at 36.9KWh (competitive), 35.9KWh (co-operative). GAVA had the most energy consumption at 67.4KWh and 65.0KWh for competitive and co-operative federated clouds respectively. 

Results of energy utilization after consolidation are shown in Fig. \ref{energy2}. From the graph, the co-operative Cloud federation model resulted in higher energy consumption compared to the competitive. This held true for all the five workload allocation schemes.

\subsubsection*{Quality of Service}
\begin{figure}[h]
\centering
\includegraphics[scale=0.5]{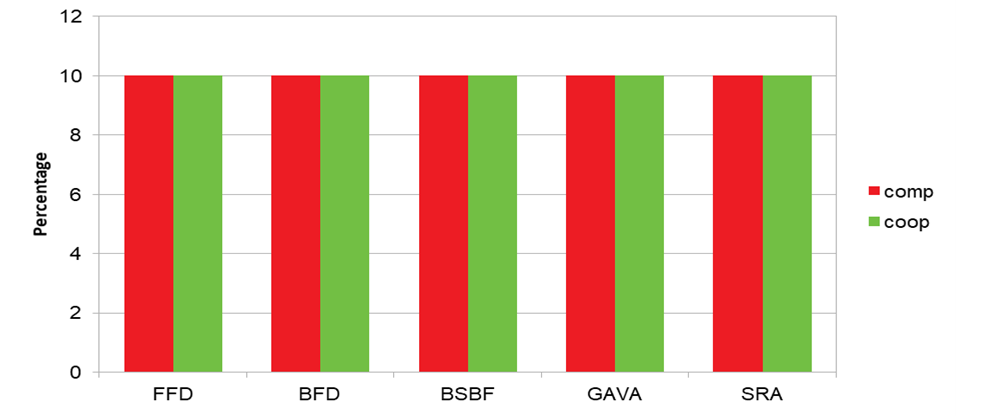}
\caption{SLA Violations due to Consolidation}
\label{slav}
\end{figure}

This is a measure of the dissatisfaction index of users to the allocation / servicing of their workloads. It is often expressed in form of a Service Level Agreement (SLA). For this work, the SLA metric used was similar to that used in \cite{BelBuy, BSBF, Hieu17} and many other works. Fig. \ref{slav}, shows a comparison of the average SLA violation for each of the algorithms and across the two cloud federation models for light user workloads. From the figure, violation percentage remained equal across all the algorithms and federation models. This result might be attributed to the fact that the users' workloads were light weight and similar and that the PMs were more than capable of serving them with minimal SLA violations.

\subsubsection*{Number of VM Migrations}
\begin{figure}[h]
\centering
\includegraphics[scale=0.5]{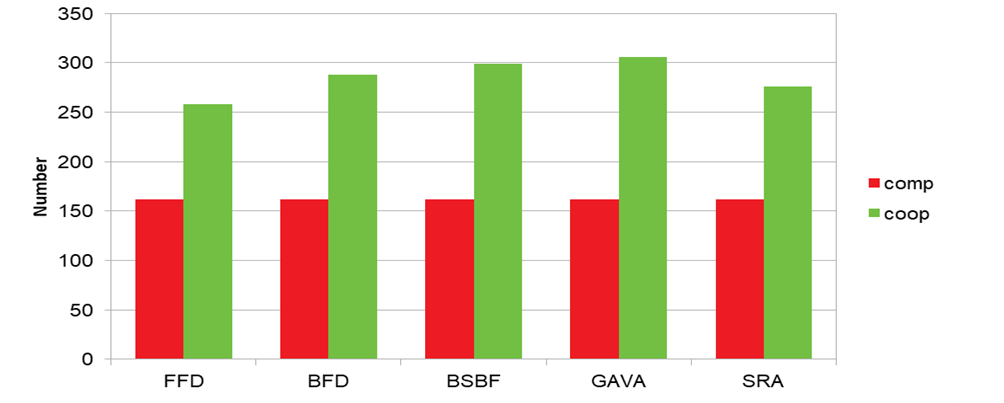}
\caption{Comparison of Number of Migrations}
\label{mig}
\end{figure}

The last metric considered is the migration count and it is a measure of the number of times, user workloads were moved to different PMs, either for consolidation purposes or to reduce SLA/QoS violations. The results in Fig. \ref{mig}, shows that user workloads are migrated more often in the co-operative federated Cloud than in the competitive. An explanation for this is that there are less resources (PMs) in the competitive than in the co-operative model, hence limited migration options. The number of migrations were equal for all allocation schemes under the competitive federation model. For the co-operative however, FFD resulted in the least number of migrations (258), followed by SRA with 276, BFD with 288, BSBF with 299 and GAVA with 306. 

\subsection*{Heavy Workloads}

\subsubsection*{Allocation Delay}

\begin{enumerate}
\item \textbf{Pre-processing Delay:}

From Fig. \ref{fallo2} and like in Fig. \ref{fallo}, BSBF again had the longest pre-processing delay for both the competitive and co-operative models at 4,611,366.67ns and 3,699,650.00ns respectively. BFD was second, at 2,905,150.00ns for competitive and 1,726,300.00ns for co-operative. FFD was the fastest of the three at 639,500.00ns for the competitive and 579,800.00ns for co-operative federation. Similar to result obtained with the light weight workloads, pre-processing delays were higher in the competitive federation model than in the co-operative model. 

\begin{figure}[h]
\centering
\includegraphics[scale=0.5]{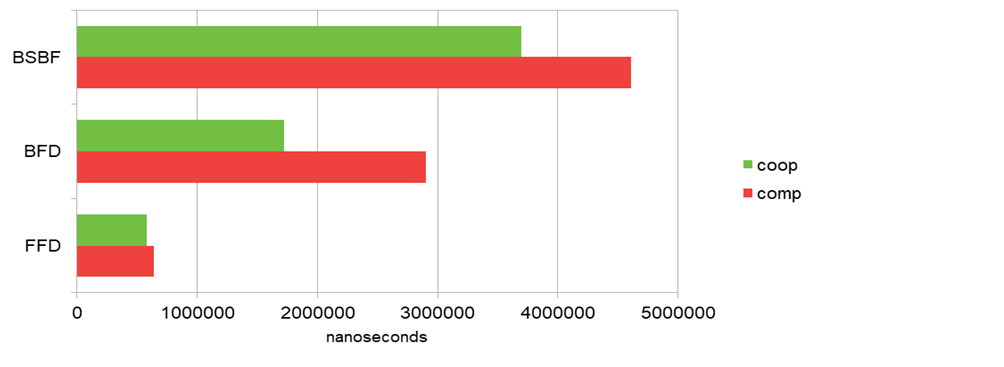}
\caption{Pre-processing Delay}
\label{fallo2}
\end{figure}

\item \textbf{Average Delay}
Fig. \ref{avg2}, shows that BFD resulted in longest delay at 756,459.00ns for co-operative and 1,299,219.00ns for competitive. FFD was the fastest at 123,338.50ns for the co-operative and 111,508.00ns for competitive. Finally, BSBF was much faster than BFD but not as fast as FFD with 191,440.67ns for the co-operative model and 194,762.00ns for the competitive. As observed with the light weight, workloads experienced lower delays in the co-operative versus the competitive Cloud federation model.

\begin{figure}[h]
\centering
\includegraphics[scale=0.5]{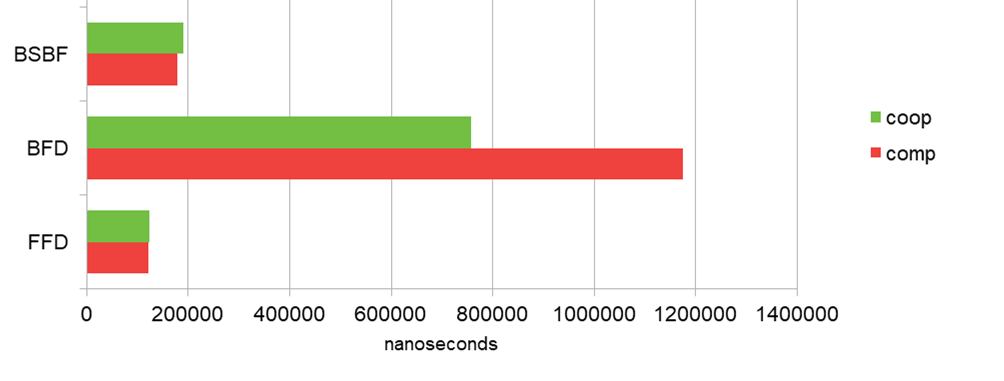}
\caption{Average workload allocation delay}
\label{avg2}
\end{figure}

\end{enumerate}

\subsubsection*{Execution Time}
\begin{figure}[h]
\centering
\includegraphics[scale=0.5]{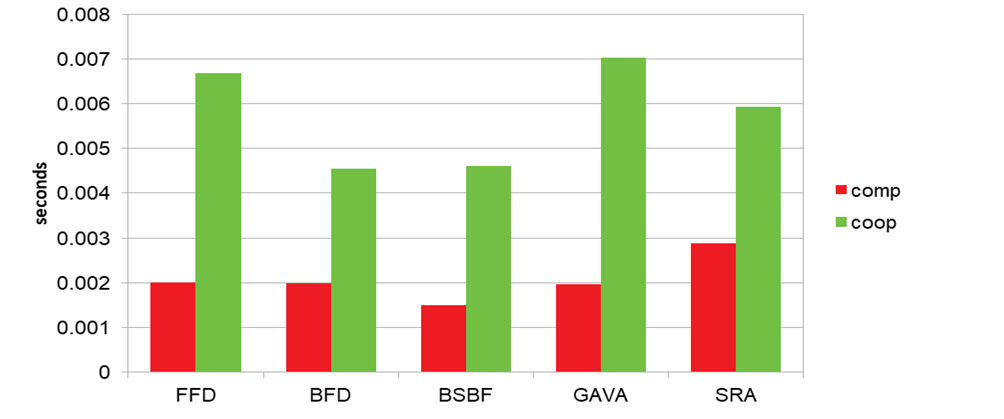}
\caption{Comparison of Overall Completion Times}
\label{time2}
\end{figure}

Fig. \ref{time2} shows a comparison of execution times of the different allocation schemes for both federation models when heavy workloads are submitted. In both the co-operative and competitive models, BSBF resulted in the quickest execution time and was closely followed by BFD. For the competitive, GAVA was the third fastest, followed by FFD and SRA; while for the co-operative federation, SRA was the third fastest, followed by FFD and GAVA. As stated above, these times difference are only in thousandth of seconds and might not be overly significant in life environments. In general and similar to the light weight workloads, execution took shorter time to complete in the competitive federation than in the co-operative federation. 

\subsubsection*{Resource Utilization}
\begin{figure}[h]
\centering
\includegraphics[scale=0.5]{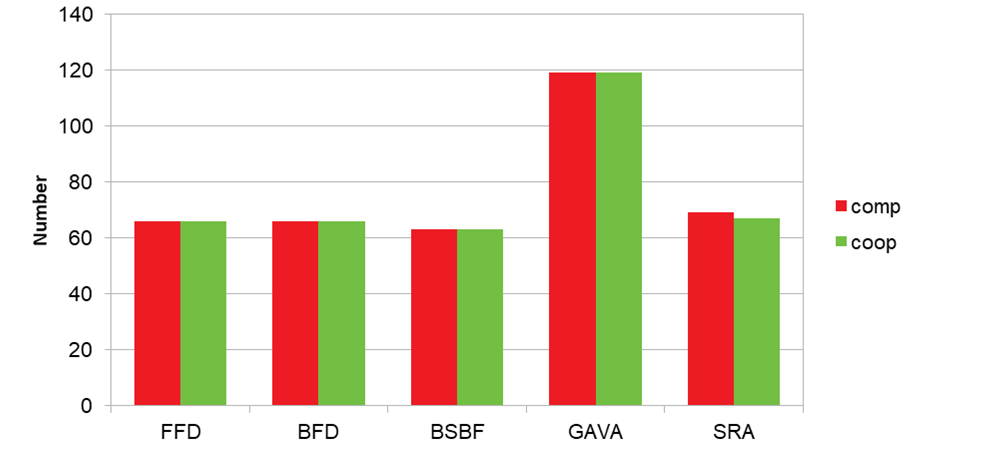}
\caption{Comparison of Resource Utilization}
\label{rub42}
\end{figure}

From the result shown in Fig. \ref{rub42}, for both co-operative and competitive federation equal number of resources were used across all but SRA. For SRA, 67 PM were used in the co-operative model compared to 69 in the competitive. Like with the light-weight workloads, across all allocation schemes the BSBF also resulted in the best matching of VMs to PMs and utilized the least number of PMs (66). BFD and FFD came second with 66 PMs, while GAVA utilized 119 PMs. comparatively, the co-operative federation model was slightly better as it utilized an average of 76.2 units of resource compared to the 76.6 used in the competitive federation.  

\begin{figure}[h]
\centering
\includegraphics[scale=0.5]{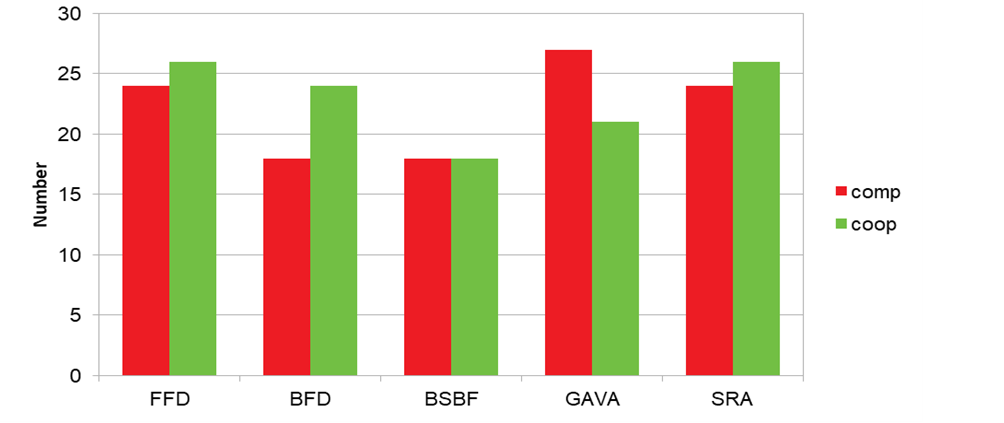}
\caption{Resource Utilization after Consolidation}
\label{ruafter2}
\end{figure}

Fig. \ref{ruafter2} shows the utilization after workload consolidation. BSBF, again resulted in the least utilization for both federation models. BFD's allocation was consolidated to 18 for the competitive and 24 for co-operative. FFD and SRA gave 24 and 26 for competitive and co-operative respectively, while GAVA resulted in 28 and 21 for competitive and co-operative respectively. Overall, resource utilization was tired across both federation models. 

\subsubsection*{Energy Conservation}
\begin{figure}[h]
\centering
\includegraphics[scale=0.5]{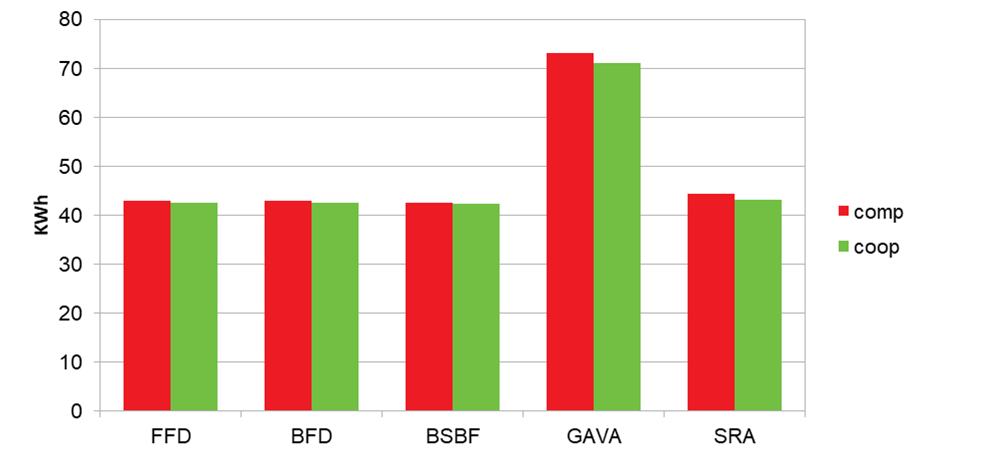}
\caption{Energy consumption before consolidation}
\label{energy12}
\end{figure}

\begin{figure}[h]
\centering
\includegraphics[scale=0.5]{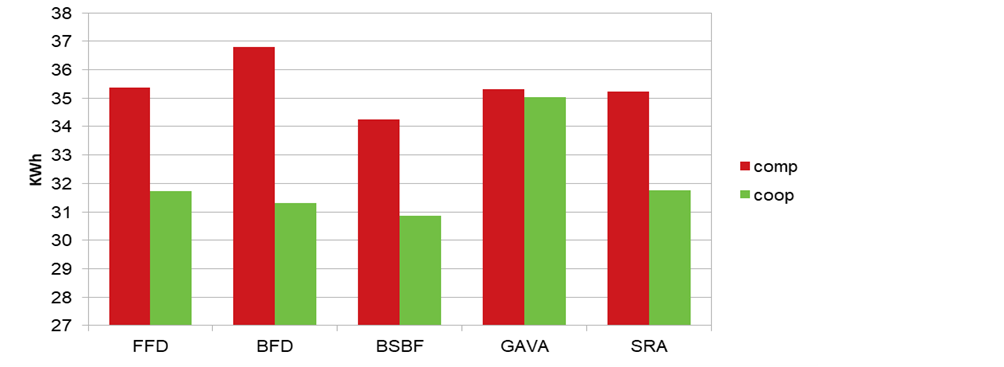}
\caption{Energy consumption after consolidation}
\label{energy22}
\end{figure}

Fig. \ref{energy12} and \ref{energy22} show a comparisons of the energy consumption when the various allocation algorithms are used to allocate heavy workloads. 

In Fig. \ref{energy12}, for the competitive federation, FFD and BFD both gave similar consumption values at 42.87KWh; while BSBF consumed 42.45KWh; SRA, 44.4KWh and GAVA, 73.14KWh. For the co-operative federation, BFD resulted in the consumption of 42.42KWh; FFD, 42.47KWh; BSBF, 42.25KWh, SRA, 43.1KWh and GAVA, 71.09KWh. Overall, less energy was consumed in the co-operative model versus the competitive model.  

Results of energy utilization after consolidation are shown in Fig. \ref{energy22}. From the graph, it can clearly be seen that for each algorithm, significantly more energy was consumed under the competitive model, than in the co-operative federation model. The only exception was GAVA where the values were closer at 35.31KWh for the competitive model and 35.03KWh in the co-operative. 

\subsubsection*{Quality of Service}
\begin{figure}[h]
\centering
\includegraphics[scale=0.5]{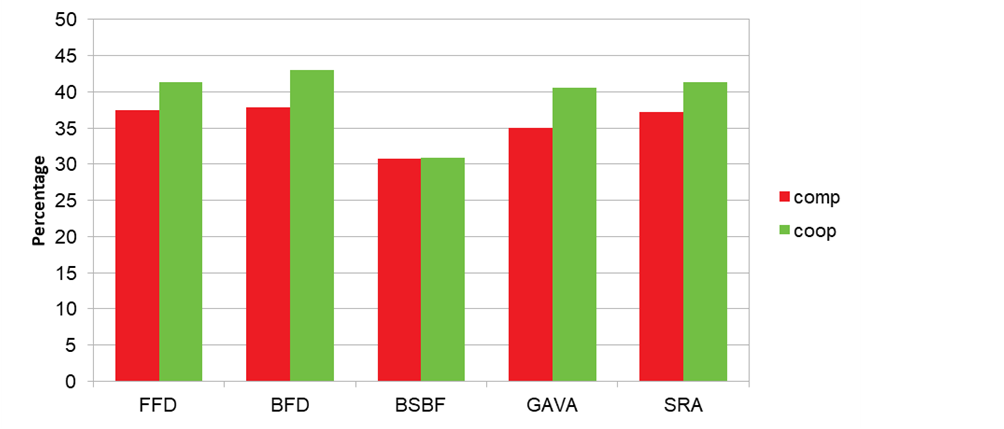}
\caption{SLA Violations due to Consolidation}
\label{slav2}
\end{figure}

Fig. \ref{slav2}, shows a comparison of the average SLA violation for each of the algorithms and across the two cloud federation models, when heavy workloads are considered. All the algorithms performed poorly, with at least 30\% workload violation. This was expected as a large proportion of the workloads required resources that could only be provided by the 2,600MHz PMs, in essence reducing the number of useable resources in the data centre by 50\%. 

BSBF resulted in the least violation of all the algorithms and for both the competitive and co-operative federation at 30.76\% and 30.85\% respectively. This was followed by GAVA with 35.05\% for competitive and 40.62\% for co-operative. The other results are as shown in the figure. Overall, workloads experienced higher violation in the co-operative federation versus in the competitive.

\subsubsection*{Number of VM Migrations}

\begin{figure}[h]
\centering
\includegraphics[scale=0.5]{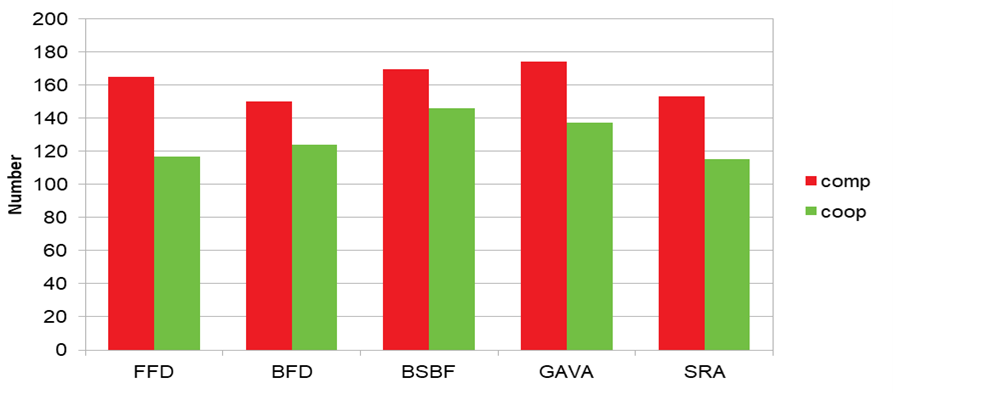}
\caption{Comparison of Number of Migrations}
\label{mig2}
\end{figure}

The results in Fig. \ref{mig2}, shows that user workloads are migrated more often in the competitive than in the co-operative federated Cloud than in the competitive.

\subsection*{Summary of Results}
Tables \ref{tab1} and \ref{tab2} provide a summary of the obtained results in a concise manner. On Table \ref{tab1}, a comparison of the two Cloud federation model is shown with their performances shown for the various metrics considered. From the table, workloads experienced lower delays in the co-operative Cloud federation but slower overall execution time compared to the competitive federation. In terms of resource utilization, the co-operative model was the better option to use for lighter workloads while competitive was best suited for heavier workloads. When energy consumption is considered, the co-operative federation is better for heavier workloads as it conserves energy better, while the competitive was better for lighter weigh workloads. In terms of providing satisfactory services, the competitive model was better overall, as it resulted in lower QoS violations for heavy workloads, while remaining at par with the co-operative federation for lighter workloads. 

\begin{table*}[t]
\begin{center}
\begin{scriptsize}
\caption{Comparison of Federation Model} \label{tab1}
\begin{tabular}{|p{2.0cm}|p{2.0cm}|p{2.0cm}|p{2.0cm}|p{2.5cm}|}
\hline
\hline
&
\multicolumn{2}{c|}{
Competitive Federation}
&
\multicolumn{2}{c|}{Co-operative Federation} 
\\
\hline
& Light Weight Workloads & Heavy Weight Workloads & Light Weight Workloads & Heavy Weight Workloads\\ 
\hline
Pre-processing Delay & High & High & Low & Low\\ 
\hline
Average Allocation Delay & High & High & Low & Low\\ 
\hline
Execution Time & Fast & Fast & Slow & Slow\\ 
\hline
Resource Utilization & Used more resources & Used more resources & Used less resources & Used less resources\\ 
\hline
Energy Conservation & High & Low & Low & High\\ 
\hline
Quality of Service & Equal & Less violations & Equal & More violations\\ 
\hline
VM Migrations & Low & High & High & Low\\  \hline 
\hline
\end{tabular}
\end{scriptsize}
\end{center}
\end{table*}

\begin{table*}[t]
\begin{center}
\begin{scriptsize}
\caption{Performance of Workload Allocation Schemes in Federation Models}
\label{tab2}
 \begin{tabular}{|p{3.0cm}|p{2.2cm}|p{3.0cm}|p{2.5cm}|p{3.0cm}|}
\hline
\hline
&
\multicolumn{2}{c|}{
Competitive Federation}
&
\multicolumn{2}{c|}{Co-operative Federation} 
\\
\hline
& Light Weight Workloads & Heavy Weight Workloads & Light Weight Workloads & Heavy Weight Workloads\\ 
\hline
Pre-processing Delay & 
\multicolumn{4}{c|}{First: FFD; Second: BFD; Third: BSBF}
 \\ 
\hline
Average Allocation Delay  & 
\multicolumn{4}{c|}{First: FFD; Second: BSBF; Third: BFD}
\\ 
\hline
Execution Time & First: GAVA; Second: SRA, BSBF, FFD; Fifth: BFD
 & First: BSBF; Second: GAVA, BFD, FFD; Fifth: SRA
 & First: BFD; Second: SRA, FFD; Fourth: BSBF; Fifth: GAVA
 & First: BFD; Second: BSBF Third: SRA; Fourth: FFD; Fifth: GAVA
\\ 
\hline
Resource Utilization (Before Consolidation) & 
\multicolumn{4}{c|} {First: BSBF; Second: BFD, FFD; Fourth: SRA; Fifth: GAVA}
\\ 
\hline
Resource Utilization (After Consolidation) & First: BSBF; Second: BFD, FFD; Fourth: SRA; Fifth: GAVA
 & First: BSBF; Second: BFD; Third: FFD, SRA; Fifth: GAVA
 & First: BSBF; Second: BFD, FFD; Fourth: SRA; Fifth: GAVA
 & First: BSBF; Second: GAVA; Third: BFD; Fourth: FFD,SRA
\\ 
\hline
Energy Conservation (Before consolidation) &  
\multicolumn{4}{c|}{First: BSBF; Second: BFD, FFD; Fourth: SRA; Fifth: GAVA}
\\ 
\hline
Energy Conservation (After consolidation) &  First: BSBF, BFD, FFD, SRA; Fifth: GAVA
 & First: BSBF; Second: GAVA; Third: SRA, FFD, BFD
 & First: BFD; Second: FFD, SRA, GAVA, BSBF
 & First: BSBF; Second: BFD; Third: FFD, SRA, GAVA
 \\ 
\hline
Quality of Service & Equal & First: BSBF; Second: GAVA; Third: SRA, FFD, BFD & Equal & First: BSBF; Second: GAVA; Third: SRA, FFD, BFD\\ 
\hline
VM Migrations & Equal & First: BFD; Second: SRA; Third: FFD; Fourth: BSBF; Fifth: GAVA
 & First: SRA; Second: FFD; Third: BFD; Fourth: BSBF; Fifth: GAVA
 & First: SRA; Second: FFD; Third: BFD; Fourth: GAVA; Fifth: BSBF
\\ 
\hline \hline
\end{tabular}
\end{scriptsize}
\end{center}
\end{table*}

Table \ref{tab2} considers the performance of the various allocation schemes across the various metrics, workload types and federation models. From the table, FFD resulted in the shortest delay, followed by BFD and BSBF. This is understandable as BFD seeks through the entire PM list for the best-fit, while BSBF has to create and constantly update its BST during the allocation process. For overall execution time, BFD was the fastest while GAVA was the slowest for the co-operative federation. For the competitive, GAVA was fastest for light weight workloads and BSBF fastest for heavy workloads. With respect to resource utilization both before and after consolidation, BSBF was the most effective and was closely followed by BFD. This can be attributed to both algorithm seeking to allocate workloads to resources that fits the best. Similar trends are observed for energy conservation (before consolidation). For energy conservation (after consolidation), BSBF was better for all but light weigh workloads in the co-operative federation where it was marginally lost to BFD.

Across both federation models, all the allocation schemes gave similar results for the light weight workloads. For the heavy workloads, however, BSBF resulted in the least violation and was followed by GAVA, SRA, FFD and BFD. Finally, in terms of VM migration, for the co-operative federation the SRA economic model resulted in the least number of migration for both heavy and light weight workloads. It performed equally well in the competitive model being only slightly outperformed by BFD. BSBF and GAVA on the other hand resulted in the highest number of migrations.

\section*{Business Models and Deployment Considerations}
\begin{table*}[h]
\begin{center}
\begin{scriptsize}
\caption{Mapping of Medical Workloads}
\label{tabMap}
 \begin{tabular}{|p{3.0cm}|p{5.0cm}|p{2.5cm}|p{2.0cm}|}
\hline \hline
Application Areas & Description & Requirements & Workload Category
\\ 
\hline
Picture Archival Communication System & Storage and communication of medical images including X-Rays, CT scans, digital pathology & IaaS providing large storage and processing system. & Heavy
\\
\hline
Medical Inventory Management System  & An information system for managing digital requisition, storage, disbursement of medical related inventories. & SaaS - software solution. & Light
\\
\hline
Online Billing System & A system for managing the hospital's billing system. & SaaS - software solution & Light
\\
\hline
Medical Training Courses & This includes learning materials, video tutorials formatted in high definition, e-books, presentation slides, lecture notes etc. & IaaS/SaaS & Heavy
\\
\hline
Patient Information System  & Digital repository of patient's medical records. & SaaS - software solution & Light
\\
\hline
Tele-medicine, Tele-surgery and collaborative surgery & Usually involves streaming very high definition audio/visuals  & IaaS/PaaS & Heavy
\\
\hline
Laboratory Management System & Specimen management and result processing system  & SaaS - software solution & Light/Heavy
\\
\hline
    Emergency and Ambulance Management Services & A system that incorporates emergency call centres, ambulance dispatch, route planning, and triage. & PaaS - a platform to incorporate multiple systems. & Light
\\ \hline \hline
\end{tabular}
\end{scriptsize}
\end{center}
\end{table*}


\begin{table*}[!ht]
\begin{center}
\caption{SWOT Analysis of Cloud Federation for Health Care in Africa}
\centering
\label{tabDep}
\begin{tabular}{|p{7cm}|p{7cm}|}
\hline \hline
Strengths & Weaknesses 
\\
\hline
\begin{itemize}
    \item Improvement in level of health care services across the African continent.
    \item Cost savings for patients 
    \item Cost savings for the hospitals and country in general
    \item Collaboration and team work between medical practitioners across Africa
    \item Potential reduction in mortality rate in developing countries across Africa
\end{itemize}
&
\begin{itemize}
    \item High cost of purchasing and installing communication facilities.
    \item There is the need to educate / train medical and support staff to use the facilities.
    \item Human inherent resistance to change.
\end{itemize}
\\
\hline
Opportunities & Threats
\\
\hline
\begin{itemize}
    \item Potential for economic growth of African countries.
    \item State-of-the-art medical and technological facilities
    \item High scalability
    \item Potential to apply machine learning and artificial intelligence to find hidden pattern and improve on medical services. 
\end{itemize}
&
\begin{itemize}
    \item Security - a security breach could result in exposure of sensitive patient information.
    \item Over-reliance on network and communication facilities - an outage or network downtime could be fatal especially in emergency situations or during a surgical procedure.
    \item Diverse polices and information usage acts across counties.
\end{itemize}
\\ \hline \hline
\end{tabular}
\end{center}
\end{table*}

\begin{enumerate}
    \item {Mapping of Medical Workloads}
    The authors in \cite{daman, wang} had presented a number of ways in which Cloud computing could be applied to medicine. Some of the application areas included: preservation of medical data, medical training, medical imagery, online billing systems, medical inventory management systems etc. These are services and/or facilities that should be available in all standard medical facilities of repute, however this is not the case for hospitals in developing countries across Africa. As stated in the introductory section of this paper, Cloud federation and collaboration can help improve the quality of medical services in Africa. To put this in perspective and tie it to the models and results presented in this paper thus far, these aforementioned Cloud medical applications can be grouped into heavy and light weight workloads based on our perception of data size and system (resource) requirements. Table \ref{tabMap} shows some potential application areas of Cloud computing in medicine and mappings to corresponding workload category.
    
    \item {Deployment Considerations}
    In considering new projects, products or process, the SWOT analysis is often used by organizations, as it easily identifies the potential weaknesses and threats. It also sheds light on the unique advantages of their product as well as the potential opportunities they can tap into. In Table \ref{tabDep}, the various aspects of the SWOT analysis of Cloud federation for health care in Africa are itemized.
          
    \item {Business Models}
    A number of business models for Cloud computing and related technologies have been discussed in \cite{Turber, Osterwalder}. This section presents business models that can be applied to Cloud federation based on a number of perspective. These models are shown on Table \ref{tabBiz}.

\end{enumerate}
\begin{table*}[!ht]
\begin{center}
\begin{scriptsize}
\caption{Potential Business Models for Cloud Federation for Health Care}
\label{tabBiz}
 \begin{tabular}{|p{1.5cm}|p{2.0cm}|p{2.0cm}|p{2.5cm}|p{2.5cm}|p{3.0cm}|}
\hline \hline
Perspective	& Federation Model & Business Model & Description & Benefits & Challenges and Limitations
\\
\hline
Between Cloud Service Providers (Business-to -Business) & Competitive Federation & Product-based: subscription model & A CSP has to pay a monthly or annual subscription fee for access to the network of other CSPs & Easy and customizable payment plans & i. CPSs are lock-in for the duration of the contract. ii. difficult to migrate date to another CSP
\\
\hline
& & Product-based: product-as- a-service model & Payment is made each time the service is used. & Vendor lock-in is avoided & Constant data migration might compromise data integrity and confidentiality.
\\
\hline
& Co-operative Federation & Service-based: Support model & The platform is made available to potential CSPs at a little entry fee, but revenue is made from service and support provided. & i.  Low entry requirements. ii. Revenue sharing model can easily be agreed upon as there are tools that can accurately measure the resource utilization of each member CSP. & i. Profitability is dependent on the amount of service provided to users. ii. Security can be a challenge, as a security breach on a member CSP can spread across the entire co-operative network. 
\\
\hline
Between Hospitals and Cloud Service Providers  & Hybrid Cloud (On-premise Cloud + Public Cloud) & Product-based: product-as- a-service model & The hospital pays each time the service is used. & If optimally used, this model can result in higher margin for the CSP and lower cost for the hospital & i.Difficult to move data to other CSPs 
ii. Patients suffer the most if there is a dispute between the hospital and CSP 
\\
\hline
Between Patients and Hospital & NA & Product-based: subscription model & Patients pay a monthly or annual subscription. The fee could be included as part of medical insurance bills. & i. Easy payment system. ii. Customizable payment options & i. Patient are unaware of the back-end Cloud provider and their QoS levels. ii. Patients have no way of ensuring that the hospital pays the CSP.
\\ \hline \hline
\end{tabular}
\end{scriptsize}
\end{center}
\end{table*}

\section*{Conclusion}
Malnutrition, epidemic diseases and high human mortality rate are a few of the common trends in many African countries. Many of these are associated with the high level of poverty and poor state of infrastructure, especially those related to health. There are however a few countries in Africa with better than average health care facilities and those with world-class ones. Thus an imbalance exist across African nations in terms of health care. A solution to this would be to build world-class hospitals across every cities across the continent, but the cost of this is prohibitively expensive. An alternative solution is to leverage on technology and Cloud computing in particular. Cloud computing has emerged as a computing paradigm that converts computing from a product to a paid service. By leveraging on the Cloud, medical expertise can be "imported" at a comparatively cheaper cost. Amongst the offering of Cloud computing is on-demand access to computing resources, cost savings. These features are however not often achievable by a single Cloud Service Provider (CSP) without adverse effect on service quality which is pertinent to health care. In a bid to achieve these without compromising quality, CSPs have to collaborate and form Cloud federations. Cloud federation across the African continent can prove to be an effective solution to some of the health care infrastructural challenges. 
In this paper, two Cloud federation models were considered - the co-operative and competitive. Five different Cloud workload allocation schemes - three heuristic based (First-Fit-Descending (FFD), Best-Fit-Descending (BFD) and Binary-Search-Best-Fit (BSBF), one meta-heuristic (Genetic Algorithm (GAVA)) and one economic model (Stable Roommate Allocation (SRA)) to determine their performance and effect on co-operative federation, where participating CSPs work and pool resources together and in competitive federation, where participants utilize their resources independently. Service delay, resource utilization, energy conservation and adherence to Service Level Agreements (SLA) were metrics considered and experimental simulations were conducted on both light and heavy workloads. 
Obtained results show that the co-operative federation resulted in the least allocation delays and utilized resources better, while the competitive federation was faster in completing user tasks with lower violations on agreed service level. With respect to the allocation algorithms, FFD was the fastest overall, while BSBF was the most effective for resources utilization, energy conservation and service adherence.
Finally, this paper presented deployment considerations for federated Cloud for health care across Africa as well as various potential business models. For future works, the effect of cost and penalties associated with SLA violations might be considered as well as a hybrid combination of these algorithms in a bid to find an optimal solution. Also for the most effective network architecture, government policies, and ethical considerations for this trans-national Cloud federation can be looked into.


\end{document}